\nonstopmode
\documentstyle[emulateapj,epsf]{article}
\makeatletter                                            
\@ifundefined{epsfbox}{\@input{epsf.sty}}{\relax}        
\def\plotone#1{\centering \leavevmode                    
\epsfxsize=\columnwidth \epsfbox{#1}}                    
\def\plotone_reduction#1#2{\centering \leavevmode        
\epsfxsize=#2\columnwidth \epsfbox{#1}}                  
\def\plottwo#1#2{\centering \leavevmode                  
\epsfxsize=.45\columnwidth \epsfbox{#1} \hfil            
\epsfxsize=.45\columnwidth \epsfbox{#2}}                 
\makeatother                                             

\def\yr{{\rm \ yr}}
\def\Myr{{\rm \ Myr}}
\def\cm{{\rm \ cm}}
\def\pc{{\rm \ pc}}
\def\G{{\rm G}}
\def\K{{\rm \ K}}
\def\kms{{\rm \ km\ s^{-1}}}
\def\cmi3{{\rm \ cm^{-3}}}

\def\simlt{\lower.5ex\hbox{$\; \buildrel < \over \sim \;$}}
\def\simgt{\lower.5ex\hbox{$\; \buildrel > \over \sim \;$}}

\begin{document}

\title{Dissipation in Compressible MHD Turbulence}
\author{James M. Stone, Eve C. Ostriker}
\affil{Department of Astronomy, University of Maryland, College Park, MD 20742-2421}
\and
\author{Charles F. Gammie}
\affil{Center for Astrophysics, MS-51, 60 Garden St., Cambridge MA 02138}

\begin{abstract}

We report results of a three dimensional, high resolution (up to
$512^3$) numerical investigation of supersonic compressible
magnetohydrodynamic turbulence.  We consider both forced and decaying
turbulence.  The model parameters are appropriate to conditions found
in Galactic molecular clouds.  We find that the dissipation time of
turbulence is of order the flow crossing time or smaller, even in the
presence of strong magnetic fields.  About half the dissipation occurs
in shocks.  Weak magnetic fields are amplified and tangled by the
turbulence, while strong fields remain well ordered.

\end{abstract}

\subjectheadings{MHD -- turbulence -- waves -- ISM: kinematics and dynamics --
ISM: magnetic fields}

\section{Introduction}

The large linewidths of molecular species in molecular clouds in our
Galaxy imply the velocity dispersion $\sigma_v$ of the gas is much
larger than the sound speed $c_s$ ($\sigma_v \sim 1-10$~km~s$^{-1}$,
whereas $c_{s} \sim 0.2-0.3$~km~s$^{-1}$; e.g. \cite{sol87},
\cite{hey97}).  Although magnetic field strengths are difficult to
measure, the best estimates in such clouds give Alfv\'en speeds
$v_{A}$ that are much larger than the sound speed but of order, or
somewhat exceeding, the velocity dispersion (e.g. \cite{mye88},
\cite{cru93}, \cite{goo94}, \cite{cru98}).  The dynamics of
this gas is of considerable astrophysical interest, as it may govern
the rate and character of star formation in our galaxy.

Two notions about the dynamics of molecular clouds have been
particularly influential since the first CO maps were made in the
1970s:  (1) The turbulent motions in molecular clouds are thought to
act as a ``turbulent pressure'' to support a cloud against self-gravity
(cf.  \cite{cha51}).  This is motivated by the discrepancy between
estimated cloud collapse times $\simlt 3\times 10^6 \yr$, after
which it is presumed that most of the cloud would turn into stars
(violating limits on the Galactic star formation rate), and estimated
cloud lifetimes $\simgt 3\times 10^7 \yr$.  (2) Supersonic,
sub-Alfv\'enic turbulence is thought to persist for more than a cloud
flow crossing time over cloud size $L$, $t_f(L)=L/\sigma_v=10^7\yr
\times (L/10\pc) (\sigma_v/1\kms)^{-1}$, because magnetic fields
provide a cushion that reduces dissipation rates.  In particular
\cite{aro75} proposed that molecular cloud turbulence may be primarily
in Alfv\'enic motions because for linear-amplitude waves no
compressions are involved.  Recently, both of these ideas have been
called into question.  This Letter describes high resolution, three
dimensional numerical experiments designed to test the latter idea,
with possible implications for the former.

We evaluate the dissipation rate of supersonic, sub-Alfvenic
turbulence in the context of an idealized numerical model.  Our model
is three-dimensional, compressible, ideal (no explicit resistivity,
viscosity, or ambipolar diffusion), non-self-gravitating (future papers
will discuss the effects of gravity), isothermal (a fair approximation
for most of the material in molecular clouds) and has a uniform
mass to flux ratio.  It is also homogeneous and isotropic, insofar as
it considers the evolution of turbulence in a periodic box, where there
are no boundaries.  

This work builds on earlier results from our group and from others.
Gammie \& Ostriker (1996; hereafter GO) considered similar issues in a
1 2/3 dimensional model, while Ostriker, Gammie, \& Stone (1998;
hereafter OGS) considered a 2 1/2 dimensional model.  Among other
results, GO found that purely Alfv\'enic turbulence quickly couples to
other, compressive waves (see \S 4 below), and OGS found that
independent of initial turbulence levels, magnetically supercritical
clouds collapse gravitationally in 5-10 Myr, in the absence of
stirring.  \cite{pad97} studied the evolution of Mach 5 decaying MHD
turbulence in 3D cloud models with $\beta=2$ and $0.02$, and
\cite{mac98} studied the evolution of Mach 5 decaying MHD turbulence
for $\beta=1$ and $0.04$; both groups concluded the dissipation time in
such models is short.  This work differs from these last two in that we
consider {\it driven} turbulence and turbulence decaying from {\it
saturated} initial conditions, thus avoiding transients associated with
a particular choice of initial conditions.

\section{Method and Parameters}

We integrate the equations of compressible, ideal MHD using the ZEUS
code (Stone \& Norman 1992a; 1992b).  The model is a cubic, periodic
box of size $L$ containing a plasma of uniform density $\rho_0$
threaded by an initially uniform magnetic field ${\bf B}_0 = (B_{0}, 0,
0)$.  The sound speed $c_s$ is constant in both space and time.  Grid
resolutions vary between $32^3$ and $512^3$.  To allow a study of
turbulent mixing, all models evolve a passive contaminant which
initially fills a cylindrical volume in the center of the grid oriented
with the symmetry axis parallel to ${\bf B}_0$ and with diameter and
axial length equal to $L/2$.

We drive turbulence by adding velocity perturbations $\delta {\bf v}$
at time intervals $\bigtriangleup t$ with $\bigtriangleup t L/c_s =
0.001$.  Each $\delta {\bf v}$ is an independent realization of a
Gaussian random field with power spectrum $|\delta {\bf v}_{k}^{2}|
\propto k^{6} \exp( -8 k/k_{pk} )$ (we choose $k_{pk} = 8 \times
(2\pi/L)$), subject to the constraint that $\nabla \cdot \delta {\bf v}
= 0$.  Since the input perturbations are incompressive, this is a
minimally dissipative way of stirring the model.  The perturbations are
normalized so that the kinetic energy input rate $\dot{E} = const.$,
and no net momentum is added to the box, $\int \rho \delta{\bf v} =
0$.

In addition to $\bigtriangleup t L/c_s$ and $k_{pk}L/2\pi$,
two dimensionless parameters characterize our models: $\beta$ and
$\dot{E}$.  We study values of $\beta = 0.01$ (strong field), 0.1
(moderate field), 1.0 (weak field), and $\infty$ (pure hydrodynamics).
Corresponding physical values of the magnetic field are given by
$B=1.4\mu\G\  \beta^{-1/2}\left({T/10\K} \right)^{1/2}
\left({n_{H_2}/10^2 \cm^{-3}} \right)^{1/2}.$ In this Letter we present
only models with $\dot{E}/ \rho_0 L^{2} c_{s}^{3} = 10^3$, but comment
on results drawn from other models.

To transform dimensionless parameters to astronomically relevant
quantities, one may independently choose values for $\rho_0$, $c_s$,
and $L$.  As an example, consider a cloud clump of size $L=2$~pc, mean
density $n_{H_2}=10^3$~cm$^{-3}$, and temperature $T=10$~K.  Then the
sound speed is $c_{s} \approx 0.2$~km~s$^{-1}$.  This implies the sound
crossing time $t_{s} \equiv L/c_s \sim 10$~Myr, and driving power
$\dot{E} = 0.4 L_{\odot}$.  For $\beta=0.01$ the magnetic field
strength is $B=44~\mu$G and the Alfv\'en speed is $v_{A} \approx
2$~km~s$^{-1}$ with the corresponding Alfv\'en wave crossing time
$t_{A} \equiv L/v_A \sim 1$~Myr.  For the moderate (weak) field case
the field strength is reduced to $14\mu\G$ ($4.4\mu\G$), so that
$v_A=0.6 (0.2)\kms$ and $t_A=3 (10)\Myr$.  A typical velocity
dispersion in a cloud with these properties is $1\kms$, so
$t_f(L)=2\Myr$.

\section{Results}

First consider driven turbulence models.  Evolution of the total energy
in fluctuations
$E \equiv E_K + E_B \equiv \int (\rho v^2/2 + (B^2 - B_0^2)/(8\pi)$ for
$\beta = 1$ models computed at resolutions of $32^3$ through $512^3$
shows that in each case, $E$ rises steeply and then reaches a final,
saturated value.  The amplitude of $E/\rho L^{3} c_{s}^{2}$ depends
quite sensitively on the numerical resolution; grids of $32^{3}$,
$64^{3}$, $128^{3}$, $256^{3}$ and $512^{3}$ zones give saturated
energy levels of 9.4, 13, 16, 17, and 18 respectively.  Note there is a
clear (although slow) trend towards convergence in these numbers,
although since we have not used identical realizations of the forcing
spectrum in each case we cannot measure the rate of convergence
precisely.  In the saturated state the dissipation rate balances the
input power $\dot{E}$, and one may define the dissipation timescales
$t_{diss} \equiv E/\dot{E}$ and $t_{diss}^K \equiv
E_{K}/\dot{E}$; these may be compared to the flow crossing time at the
scale $\lambda_{pk}=L/8$ at which the turbulence is driven,
$t_f(\lambda_{pk}) \equiv \lambda_{pk}/\sqrt{2 E_K}$ (which we
hereafter abbreviate as $t_f$).\footnote{$t_f$ is often referred to as
the ``eddy turnover time'' for incompressible turbulence; for
compressible flows the present terminology is preferred.} All the
models saturate at times $\sim t_f$.  For the $\beta=1$ model at
$512^3$ resolution, we find $t_{diss}/t_f = 0.75$ and $t_{diss}^K/t_f
= 0.54$.

Where does the energy go in the numerical models?  One route to
dissipation is via shocks.  ZEUS uses an artificial viscosity to
capture shocks, and so the shock dissipation rate can be measured by
integrating the work done by artificial viscosity over space and time.
This accounts for about $50\%$ of the dissipation.  Another route is
through a turbulent cascade like that which occurs in incompressible
hydrodynamic (e.g. \cite{lan87}) and MHD (e.g.  \cite{gol95})
turbulence; there nonlinear interactions transfer energy to
progressively smaller and smaller scales until a dissipation scale is
reached and the energy is thermalized.  Since the present models
include no explicit resistivity, viscosity, or ambipolar diffusion to
thermalize the energy at small scales, energy is finally lost through
numerical effects at the grid scale.\footnote{Explicitly resistive
  experiments done by us capture another $20\%$ of the dissipation.
  The relative importance of the different routes to dissipation at
  small scales will depend on the precise values of the microscopic
  diffusion coefficients (e.g. \cite{bw90})} A completely satisfactory
study would include astrophysically appropriate values for the
microscopic diffusion coefficients and close the energy equation, but
we have not done so here because the task is prohibitively expensive.
This is one of the major challenges for future work.

How does the dissipation rate vary with magnetic field strength?
Figure 1a shows the evolution of $E$ for various $\beta$ at a
resolution of $256^{3}$.  The amplitude of $E$ increases
monotonically with field strength (decreasing $\beta$); hence
dissipation decreases as field strength increases.  In Table 1, we give
the values for the energy in the saturated state (averaged over time
$t=0.2-0.3t_s$), as well as the saturated-state dissipation times for
the four models displayed.  From the values in the Table, the change in
$E$ with $\beta$ is not large, amounting to only a $\sim 30\%$
increase in the $E$ saturation amplitude as $\beta$ varies from
$\infty$ to $0.01$.  The dissipation times for saturated turbulence all
lie in the range $\sim 0.5-0.8 t_f$, with slightly longer dissipation
times for stronger-$B_0$ models.

\begin{deluxetable}{crccrcccc}

\small
\tablecaption{Dissipation Characteristics of Saturated MHD Turbulence\label{tbl-1}}
\tablewidth{6.5in}
\tablehead{
\colhead{model}
& \colhead{$\beta$} 
& \colhead{$E/\rho L^{3} c_{s}^{2}$} 
& \colhead{$E_K/\rho L^{3} c_{s}^{2}$} 
& \colhead{${\delta E_B\over E_K}$}
& \colhead{$t_{diss}\over t_f$\tablenotemark{a}} 
& \colhead{$t_{diss}^K\over t_f$\tablenotemark{a}} 
& \colhead{$t_{dec}\over t_f$\tablenotemark{a}} 
& \colhead{$t_{dec}^K\over t_f$\tablenotemark{a}} 
}

\startdata
 A& $0.01$& 20.3& 13.0& 0.56& 0.83& 0.54& 0.82& 0.65\nl 
 B&  $0.1$& 18.9& 11.8& 0.61& 0.74& 0.46& 0.69& 0.39\nl
 C&  $1.0$& 17.0& 12.9& 0.32& 0.70& 0.53& 0.58& 0.37\nl
 D&  $\infty$& 15.4& 15.4& 0& 0.69& 0.69& 0.40& 0.40\nl
\enddata
\tablenotetext{a}{$t_{diss}$, $t_{diss}^K$, $t_f$, $t_{dec}$, and $t_{dec}^K$ defined in text}
\end{deluxetable}

The structure of driven compressible MHD turbulence changes as the
field strength is varied.  Figure 2 shows the logarithm of the density
along three faces of the computational volume, representative magnetic
field lines, and an isosurface of the passive contaminant after
saturation for both $\beta=0.01$ and $\beta=1$ models computed at
a resolution of $256^{3}$.  In both cases, the density is compressed
into small scale knots and filaments; in the $\beta=0.01$ model these
are elongated in the direction parallel to the field.  The mass
(volume) weighted mean of $\log(\rho/\rho_0)$ in the strong magnetic
field model is 0.28 (-0.29), whereas for the weak field model it is
0.20 (-0.22), indicating the density contrasts are larger for strong
fields at fixed turbulent Mach number.  The maximum density in the
strong field model is 83; for the weak field model it is 44.
The passive contaminant is confined to a narrow range of flux tubes for
$\beta=0.01$, indicating that cross-field diffusion is small; for
$\beta=1$ it diffuses isotropically.

There is a tendency toward equipartition of kinetic and magnetic
energy in all the models.  From Table 1, the turbulent magnetic energy
$\delta E_B$ is between 30\%-60\% of $E_K$.  In the weak field case,
significant amplification of the magnetic field is produced by the
turbulence, so that after saturation the energy in the fluctuations in
the field is ten times larger than that in the mean field.  In the
weakly magnetized model the field lines are thoroughly tangled
(Fig. 2b).  In the strong field model the field lines are relatively
well ordered (Fig. 2a), as expected (e.g. \cite{wei66}).  

Next consider models of decaying turbulence.  The initial conditions
are taken from the saturated driven models presented above.  Figure 1b
shows the evolution of $E$ for decay from saturated initial conditions
for various magnetic field strengths.  At late times the decay of $E$
follows a power law in time, with index between 0.8-0.9 (consistent
with the finding of \cite{mac98}).  This implies that the dissipation
time varies with time.  We define decay times $t_{dec}$ ($t_{dec}^K$)
as the time taken for 50\% of the {\em initial} energy (kinetic
energy) to be lost; values for the decay time in these decay runs are
given in Table 1.  For all models, the decay times are in the range
0.4-0.8 $t_f$, comparable to the range of steady-state dissipation
times.

The decay rate measured here could in principle differ substantially
from decay simulations that begin with unsaturated initial conditions.
To investigate this possibility, we have computed the decay of
supersonic turbulence from initial conditions in which the magnetic and
velocity field perturbations are taken from the saturated, driven model
A, but the density is reset to a uniform value.  The result is plotted
as a dashed line in Figure 1b.  The corresponding decay times are
$t_{dec}/t_f = 0.80$ and $t_{dec}^K/t_f = 0.68$, nearly identical to
those for Model A's decay.  

Finally, to make contact with other studies of decaying MHD turbulence,
we have performed simulations which begin with a uniform density and
magnetic field, and velocity perturbations that follow a $k^{-2}$
spectrum normalized to have the same initial energy as our driven
turbulence simulations at saturation.  The result is shown as a dotted
line in Figure 1b; the decay times for this model are $t_{dec}/t_f =
1.0$ and  $t_{dec}^K/t_f = 0.6$, again comparable to the other
dissipation times we have found.  Thus we conclude that turbulent decay
times are not strongly affected by specifics of initial conditions.
The energy decay times found for 2.5D models (OGS) are 
a factor 1.5-1.75 times larger than those obtained here with 3D models.

\section{Discussion}

Taking together the results of all the models presented here, our
conclusion is that compressible MHD turbulence dissipates rapidly -- in
less than one flow crossing time at the energy-containing scale --
regardless of the field strength, and of the details of the initial or
ongoing energy input characteristics.  For our model $2\pc$ cloud, the
turbulent dissipation time is always less than 1/4~Myr when the
turbulent scale is 1/8 the size of the cloud, and less than 2 Myr when
the largest turbulence scale is the same size as the cloud.  For GMCs,
with flow crossing times $\sim 5-10\Myr$ and the largest
energy-containing scale probably comparable to the cloud scale, our
results imply that without continual energy input, the observed
nonthermal linewidths would decay in less than the cloud lifetimes.
The energy input rate required to keep the turbulence going is not
large, however, amounting to only $0.4~\L_{\odot}$ in mechanical power,
input at a scale $0.25\pc$, for the parameters of our model cloud.

Why are magnetic fields unable to reduce dissipation in supersonic,
sub-Alfv\'enic turbulence?  One might have expected that nonlinear but
still incompressive Alfv\'en waves could safely store a significant
fraction of the energy.  But for the nonlinear amplitudes $\delta
v_A\simgt 0.1 v_A$ and $\beta\equiv c_s^2/v_A^2<1$ conditions which are
likely to obtain in molecular clouds, couplings between the MHD wave
families are strong.  Even a circularly polarized Alfv\'en wave, which
is an exact nonlinear solution to the equations of motion, is
dynamically unstable to decay into compressive motions (\cite{sag69},
\cite{gol78}).  Thus a spectrum of Alfv\'en waves of nonlinear
amplitude can be quickly converted to compressive motions, which decay
rapidly.

The dissipation of turbulent energy may be an important heat source
within molecular clouds.  From our driven-turbulence models, where
energy is input at 1/8 the size of the cloud, we find that the volume
dissipation rate of energy is $\dot E/L^3 =7.5 \rho_0 \sigma_v^3/L$;
for clouds where the energy is primarily contained on the largest scale
possible, the dissipation rate might be reduced by a factor 8 (for a
given velocity dispersion).  The corresponding volume-averaged heating
rate is
$\Gamma_{turb}=5.8\times 10^{-27} ({n_H/1\cm^{-3}})
\left({\sigma_v/\kms}\right)^3 \left({L/\pc}\right)^{-1}
{\rm erg} \cm^{-3}{\rm s}^{-1}$
for the smaller-scale turbulence case, with a factor 8 reduction
possible.  Except for conditions where the velocity dispersion
approaches the sound speed (e.g. dense cores of size $\sim 0.1\pc$, cf.
\cite{goo98}), this {\it average} turbulent heating rate exceeds the
cosmic-ray heating rate, and in large clouds where the velocity
dispersion is large, $\Gamma_{turb}$ may compete with photoelectric
heating (e.g. \cite{spi78}).  Locally, however, turbulent dissipation
may greatly exceed other sources of heating; future studies will
characterize the localization of turbulent heating.

\acknowledgements

This work is supported in part by NASA grant NAG 53840


\begin{figure}[h]
\plottwo{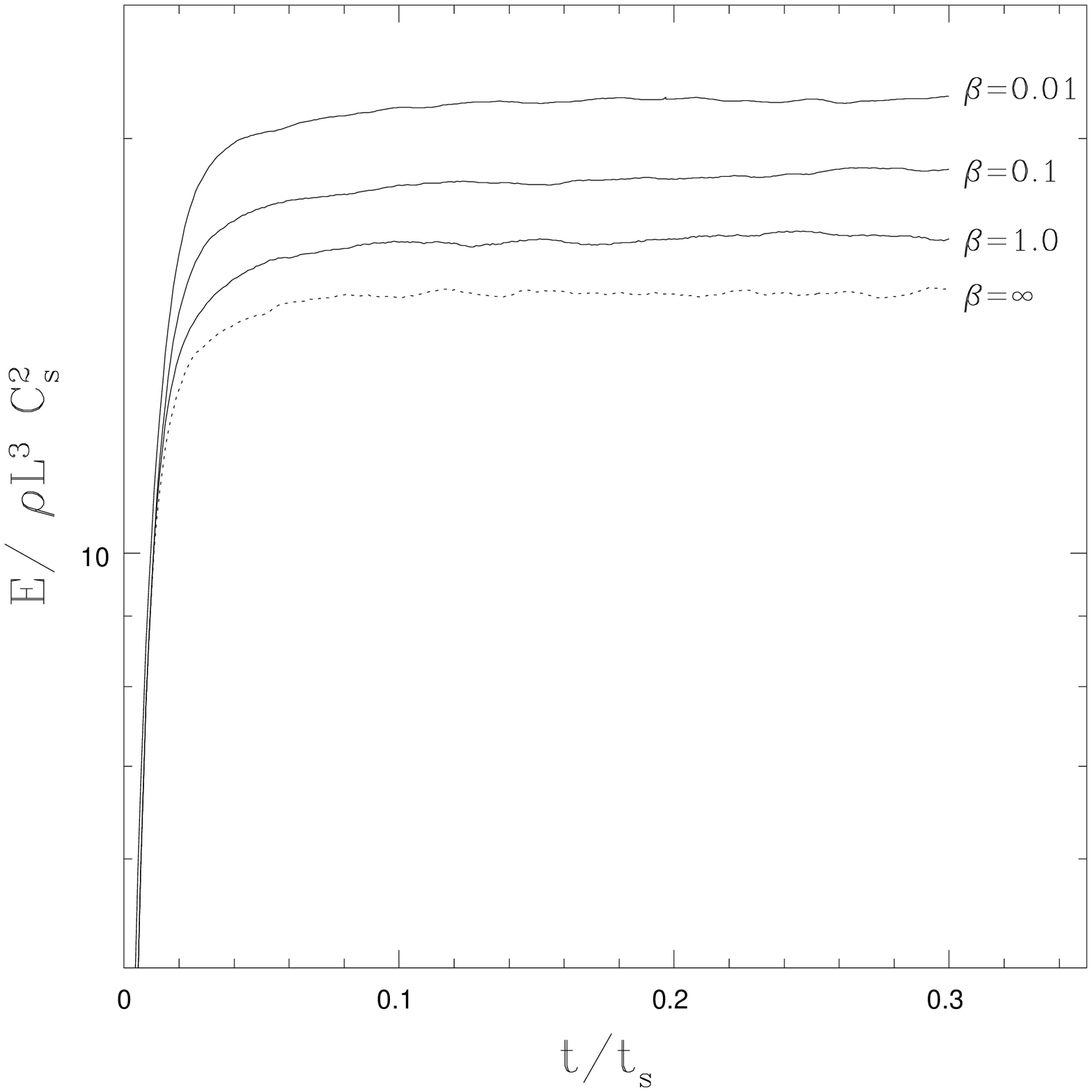}{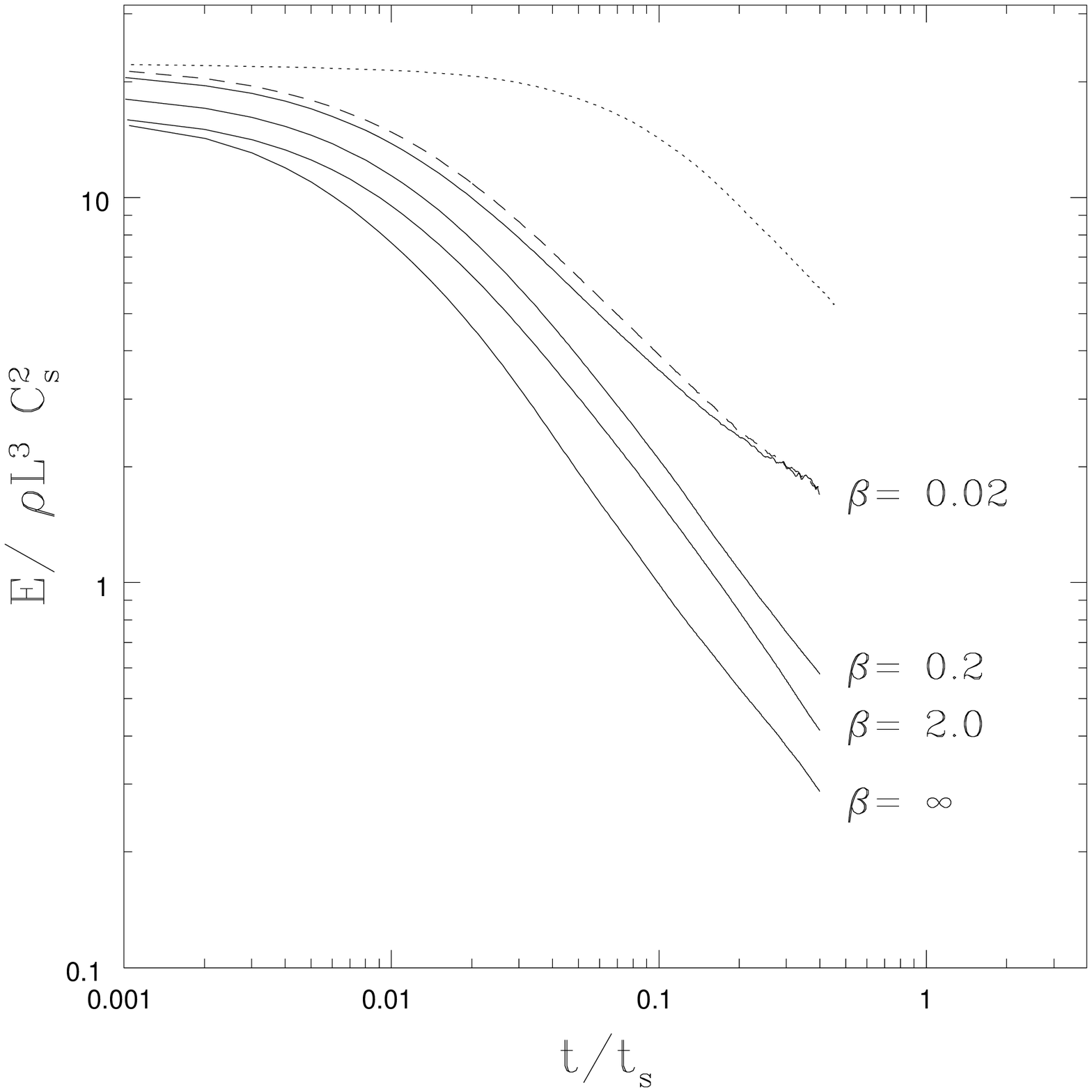}
\caption{Total energy in fluctuations $E$ versus time for various $\beta$ at a numerical resolution
of $256^{3}$  {\it Left:} forced models {\it Right:} decaying models.
Solid curves -- decay from saturated state; dashed curve -- decay of 
density-reset saturated state; dotted line -- decay of $k^{-2}$ velocity
spectrum}
\end{figure}

\clearpage

\begin{figure}
\caption{Images of the logarithm of the density (colors) on three faces of
the computational volume, representative magnetic field lines (dark blue 
lines), and isosurface of the passive contaminant (red) after saturation.
{\em Left:} $\beta=0.01$  {\em Right:} $\beta=1$}
\end{figure}

\clearpage

\end{document}